\documentclass[10pt, conference, compsocconf]{IEEEtran}
\ifCLASSINFOpdf
\else
\fi
\usepackage{cite}
\usepackage{multicol}
\usepackage{verbatim}
\usepackage{amsmath,amssymb,amsfonts}
\usepackage{algorithmic}
\usepackage[linesnumbered,ruled,vlined]{algorithm2e}
\SetCommentSty{mycommfont}
\usepackage{graphicx}
\usepackage{amsthm}
\newtheorem{definition}{Definition}
\graphicspath{{./figures/}}
\usepackage{textcomp}
\usepackage{xcolor}
\usepackage{subfig}
\begin{document}
	\title{Efficient On-Chip Multicast Routing based on Dynamic Partition Merging\\}
	
	\author{\IEEEauthorblockN{Binayak Tiwari, Mei Yang, Yingtao Jiang}
		\IEEEauthorblockA{Department of Electrical and Computer Engineering\\
			University of Nevada Las Vegas\\
			Las Vegas, USA\\
			Email: btiwari@unlv.nevada.edu, \{mei.yang, yingtao.jiang\}@unlv.edu}
		\and
		\IEEEauthorblockN{Xiaohang Wang}
		\IEEEauthorblockA{School of Software Engineering\\
			South China University of Technology\\
			Guangzhou, China\\
			Email: xiaohangwang@scut.edu.cn}
	}
	
	
	%Telephone: (800) 555--1212, Fax: (888) 555--1212}


	\maketitle
	
	\begin{abstract}
		Networks-on-chips (NoCs) have become the mainstream communication infrastructure for chip multiprocessors (CMPs) and many-core systems. The commonly used parallel applications and emerging machine learning-based applications involve a significant amount of collective communication patterns. In CMP applications, multicast is widely used in multithreaded programs and protocols for barrier/clock synchronization and cache coherence. Multicast routing plays an important role on the system performance of a CMP. Existing partition-based multicast routing algorithms all use static destination set partition strategy which lacks the global view of path optimization. In this paper, we propose an efficient Dynamic Partition Merging (DPM)-based multicast routing algorithm. The proposed algorithm divides the multicast destination set into partitions dynamically by comparing the routing cost of different partition merging options and selecting the merged partitions with lower cost. The simulation results of synthetic traffic and PARSEC benchmark applications confirm that the proposed algorithm outperforms the existing path-based routing algorithms. The proposed algorithm is able to improve up to 23\% in average packet latency and 14\% in power consumption against the existing multipath routing algorithm when tested in PARSEC benchmark workloads.    
	\end{abstract}
	

	% \fi
	\IEEEpeerreviewmaketitle

	\section{Introduction}
	Rapid advancement in semiconductor technology has made it possible to integrate tens, hundreds and even thousands of processing elements on a single chip multi-processor (CMP). Networks-on-Chips (NoCs) \cite{b21} have become the mainstream intra-chip communication infrastructure for CMPs and many-core systems due to their scalability and power efficiency. The communication in NoCs can be categorized into unicast (one-to-one) and multicast (one-to-many) communication. A significant amount of multicast traffic is demonstrated in parallel applications involving multithreaded programs, replication, barrier/clock synchronization \cite{b2}, and cache coherence \cite{b1} protocols. Simulations of a set of CMP benchmark applications with MESI cache protocols show that the multicast traffic percentage ranges from $5\%-15\%$ \cite{b23}. The study in \cite{b1} shows that depending on the coherence protocol used, the multicast traffic percentage can vary with up to $16$ destinations. Emerging heterogeneous CPU-GPU many-core systems specifically target to machine learning applications which contain large amount of communication (around $80\%$) to the memory controller \cite{b3}. Convolution Neural Network (CNN) and Deep Neural Network (DNN) \cite{b3}\cite{b4}\cite{b22} are commonly used deep learning architectures. CNNs contain a stack of convolution and pooling layers followed by a fully-connected layer at the end. The convolution and pooling layers use distributed operations for moving data between the cores, while the fully connected layer is a communication intensive process where the traffic is inherently multicast in nature. All these applications require attention for the multicast traffic.
	
	Although multicast communication can be performed by splitting a multicast packet into multiple unicast packets, it is not an efficient way of doing it due to poor resource utilization. Authors in \cite{b1}\cite{b23} show that a small percentage of multicast traffic can significantly degrade the network's performance. Thus, an efficient multicast algorithm is needed to overcome this degradation. Multicast routing algorithms are broadly classified into tree-based and path-based. In tree-based \cite{b8} algorithms, a spanning tree is constructed from the source node to cover all destination nodes. In a tree, messages are replicated at the intermediate node and forwarded along multiple branches. Although the tree-based algorithm can deliver messages to each destination via the shortest path, it is susceptible to deadlocks because of high blocking probability \cite{b5}. The multiple unicast routing can be considered as a simple tree-based algorithm which has an advantage when the multicast destination is sparsely located.
	
	In path-based multicast algorithms, intermediate branching is prohibited. Multiple paths at the source node are identified and the packet is delivered to the destination nodes along the paths. The issue with deadlock is not present in path-based algorithms, however, the path to each destination tends to be long resulting in longer packet latency. In the dual-path \cite{b6} algorithm, the destination set is divided into two partitions $D_H$ and $D_L$ covering all the destination nodes with higher and lower label number than the source node, respectively. Then the multicast packet is delivered along $D_H$ and $D_L$ paths. In the multi-path (MP) \cite{b7} algorithm, $D_H$ and $D_L$ are further divided into two subsets, one enclosing destinations with $x$ coordinate less than that of the source node ($D_{H1}$ and $D_{L1}$) and the other having destinations with $x$ coordinate greater or equal than that of the source node ($D_{H2}$ and $D_{L2}$). Each multicast packet is distributed on four paths to four destination groups. By making full utilization of four outgoing channels at the source node, the multi-path algorithm achieves better performance than the dual-path algorithm. However, the existing multi-path algorithms all use static partitions which lack the global view of path optimization.
	
	In this paper, we first formulate the multicast destination set partition problem and propose an efficient dynamic partition merging (DPM) algorithm to solve it. The proposed algorithm divides the multicast destination set into partitions dynamically by comparing the routing cost of different partition merging options and selecting the merged partitions with lower cost. The multicast routing in each partition takes advantage of both multiple unicast and dual-path algorithms. Deadlock is avoided by separating the physical network in high and low-channel networks with corresponding turn restrictions. Simulation results of the proposed algorithm by both synthetic traffic and PARSEC benchmark workloads show that the DPM-based multicast routing algorithm achieves significant improvement over the existing multi-path algorithms in both average packet latency and power consumption.
	
	The rest of the paper is organized as follows. Section \ref{relatedwork} provides brief review of existing on-chip multicast routing algorithms. In Section \ref{proposed}, the problem of multicast destination set partition is formulated and the proposed algorithm is described in detail. Section \ref{experiment} provides the experimental settings, simulation results, and discussion of the results. Section \ref{conclusion} concludes the paper.
	
	\section{Related Work} \label{relatedwork}
	In the literature, several tree-based and path-based multicast routing algorithms have been proposed for NoCs. In the Virtual Circuit Tree Multicast (VCTM) \cite{b1} scheme, a unique VCT ID is created for each source destination set and a virtual circuit is formed before the multicast traffic begins to flow. The overhead of virtual circuit setup time and tree maintenance information is not trivial, especially when the number of multicast destinations is high. For Recursive Partitioning Multicast (RPM) \cite{b9}, the network is first partitioned according to the source node location and selectively a tree is formed by replication. After the replica reaches the next hop in each partition, the same process repeats recursively inside the partition till the message is delivered to all destinations. In \cite{b25}, a deadlock free tree-based multicast routing based on a hold-release tagging method is proposed where flits can be interleaved and pheromone tracking method is used to reduce the communication energy. 
	
	Two multicast routing algorithms are proposed for 3D NoCs \cite{b29}, where MXYZ is a tree-based dimension order routing algorithm where packets are replicated to X, Y or Z direction based on the destinations with common coordinates and AL+XYZ finds an alternate output channel if the selected output channel is not available. In \cite{b13}, an adaptive and deadlock free tree-based multicast routing scheme is proposed by interleaving flits from different sources in a message queue. The K-Means based Multicast Routing (KMCR) scheme \cite{b10} employs K-means clustering method to partition the network balanced in terms of destinations. Then the centroid of each partition is found, a tree from the source node to each centroid is created and finally a subtree is formed from the centroid to each destination. All tree-based routing algorithms require the support of flit replication in the router microarchitecture.
	
	In dual-path (DP) \cite{b6} and multipath (MP) \cite{b7} routing algorithms, the destination nodes are partitioned into two groups and four groups respectively according to their relative position to the source node. In the column-path \cite{b12} based method, the message is traversed on X dimension to the first node which x coordinate is matched with the destination along the East or West direction and then delivered along the North or South direction to match the y coordinate of the destination. Authors in \cite{b11} have proposed a low distance multipath algorithm wherein the destination list is sorted according to hop count instead of the label. The algorithm is shown to be a better option than both multipath and column-path algorithms. In \cite{b14}, a path-based adaptive routing scheme is proposed to exploit potential alternative paths by prohibiting the turns, which achieves better performance with a high degree of adaptivity than \cite{b15}. 
	
	To reduce the latency in multicast packets, the path-based routing algorithm proposed in \cite{b27} selects the available output channel according to the size of buffer space in downstream routers based on the destination partitions. Authors in \cite{b31} proposed a balanced partitioning method in a 3D NoC which is an extended version of dualpath, multipath and column-path method. Authors in \cite{b32} also proposes a hierarchical multicast routing algorithm combining both multiple unicast and multi-path routing which shows better resilience to hardware trojan attacks to multicast traffic. All the aforementioned path-based multicast routing algorithms use a static destination set partition strategy which lacks the global view of path optimization. As a matter of fact, the partition strategy should be flexible enough to cope with various distribution of destination nodes.
	
	\begin{figure}
		\centering
		\includegraphics[width=1\linewidth]{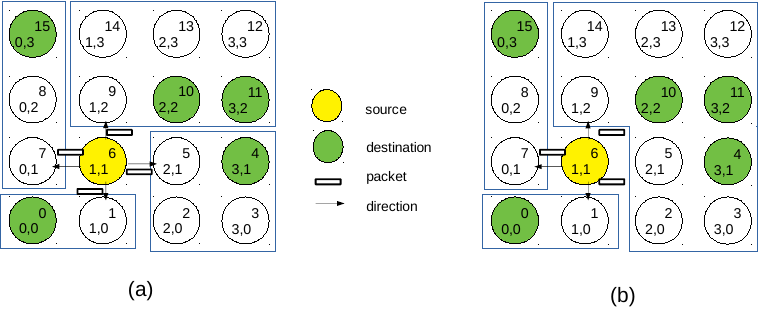}
		\caption{View with (a) Static Partition (b) Alternate Partition}
		\label{fig-globalview}
	\end{figure}
	
	\begin{figure}
		\centering
		\includegraphics[width=1\linewidth]{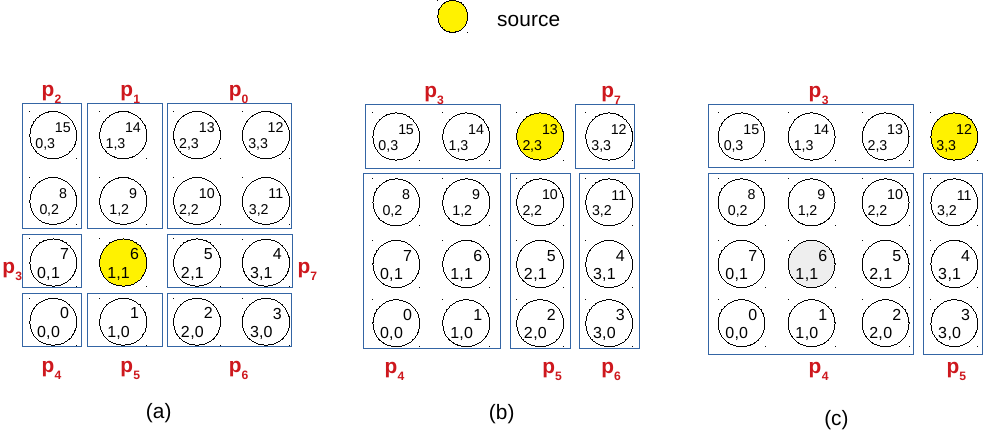}
		\caption{Partition with source node S}
		\label{fig-partition}
	\end{figure}
	
	\section{Dynamic Partition Merging-based Multicast Routing} \label{proposed}
	The problem of static destination set partition is illustrated in Fig. \ref{fig-globalview}. Based on the source node, the destination set is divided into four partitions using the partition method proposed in \cite{b7}. Thus, four packets will be sent along the four paths as shown in Fig. \ref{fig-globalview}a towards four destination groups. 
	Fig. \ref{fig-globalview}b shows an alternate partition strategy which yields less total hop count from the source node to all destination nodes than the static four-partition strategy. This example motivates us to study the problem of an optimal partition of the destination set such that the total hop count of the multicast routing paths is minimized. In the following, we first formulate the problem and propose the Dynamic Partition Merging (DPM)-based multicast routing algorithm.
	
	\subsection{Problem Formulation} \label{problemformulation}
	The destination set partition problem can be modeled as follows. Given a source node $S$ and a universal set of the $N$ multicast destination nodes $D = \{d_0,d_1,...,d_{N-1}\}$. A collection of subsets of $D$, denoted as $V=\{V_0,V_1,...,V_{k-1}\}$ where $V_0,V_1,...,V_{k-1} \subseteq D$ can be formed with the cost $C_0,C_1,...,C_{k-1}$ correspondingly, where $C_i$ is the cost of multicast routing from source $S$ to $V_i$. The objective of the problem is to find the set $I \subseteq V$ with the minimum total cost such that all destination nodes in $D$ are covered, i.e.,\\
	
	\begin{equation}
	minimize\ \sum_{V_i \in I}C_i \nonumber
	\end{equation}
	
	\textit{subject\ to:} 
	
	\begin{equation} \label{eq1} 
	\bigcup_{V_i \in I} V_i = D \end{equation}
	\begin{equation} \label{eq2} \forall V_i,V_j \in I\ and\ i\neq j, V_i \cap V_j = \emptyset \end{equation}
	
	Equation (\ref{eq1}) ensures that all the multicast destination nodes are reached. Equation (\ref{eq2}) ensures that each destination is reached only once. The destination set partition problem is actually an exact weighted set cover problem \cite{b30} which is proven to be NP-complete.
	
	\subsection{Proposed Solution}
	Considering the hardware implementation cost, we propose a heuristic algorithm to solve the destination set partition problem. In light of the fine-grained partition used in \cite{b9}, the destination set is partitioned into at most eight partitions according to the relative position to the source node $S=(s_x,s_y)$, as shown in Fig. \ref{fig-partition}a. For each non-corner edge node Fig. \ref{fig-partition}b, there are five partitions. While for each corner node Fig. \ref{fig-partition}c, three partitions are possible. Without loss of generality, we describe the DPM algorithm using a non-edge node. For each node $L = (l_x,l_y)$, its basic partition is obtained according to the following rule:
	
	\begin{itemize}
		\item $P_0=\{L:l_x>s_x\ \&\ l_y>s_y\}$
		\item $P_1=\{L:l_x=s_x\ \&\ l_y>s_y\}$
		\item $P_2=\{L:l_x<s_x\ \&\ l_y>s_y\}$
		\item $P_3=\{L:l_x<s_x\ \&\ l_y=s_y\}$
		\item $P_4=\{L:l_x<s_x\ \&\ l_y<s_y\}$
		\item $P_5=\{L:l_x=s_x\ \&\ l_y<s_y\}$
		\item $P_6=\{L:l_x>s_x\ \&\ l_y<s_y\}$
		\item $P_7=\{L:l_x>s_x\ \&\ l_y=s_y\}$
	\end{itemize}
	
	A basic destination partition set $P = \{P_0,P_1,..,P_7\}$ is formed where $P_i \subseteq D$ and all the elements of $P_i$ are mutually exclusive. An extended partition set is defined as $\mathbb{P} = \{P_0P_1,P_1P_2,..,P_7P_0,P_0P_1P_2,P_1P_2P_3,..,P_7P_0P_1\}$ where $P_iP_j$ stands for $P_i\cup P_j$ and $P_iP_jP_k$ stands for $P_i\cup P_j\cup P_k$ for $i,j,k=0,1,...7$. $\mathbb{P}$ is obtained by merging up to 3 consecutive partitions from the partition set $P$. The limit is set to 3 in order to prevent the final partition from being converged to two partitions like in the dual-path algorithm\cite{b6} which is already proven to be worse than the multipath algorithm \cite{b6}. Additionally, this limit will also help in saving unnecessary computation effort. The collection set $V=P\cup \mathbb{P}$ which contains all possible partitions by merging up to 3 consecutive partitions. The cost $C_i$ of multicast routing for each destination partition in set $V$ is obtained as defined in Definition \ref{definition-treevspath} which is determined after finding the representative node (Definition \ref{definition-representative}) in the partition. The saving $A$ defined in Definition \ref{definition-saving} is used to measure the merging benefit for each element in $\mathbb{P}$.

	\begin{definition} \label{definition-representative}
		The $representative\ node$ $R = (r_x,r_y)$ \textup{of an element} $V_i \in V$ \textup{is the nearest destination node} $d = (d_x,d_y)$ \textup{from a source node} $S = (s_x,s_y)$ \textup{for a given multicast packet obtained by:} R = $\underset{d}{arg\ min} \ \{cost_{d} : \forall \ d \ \epsilon \ V_i\}$ , \textup{where} $cost_d = |d_x-s_x|+|d_y-s_y|$.
	\end{definition}
	
	\begin{definition} \label{definition-treevspath}
		
		\textup{The} $cost$ $C_i$ \textup{for each destination partition} $V_i \in V$ \textup{can be obtained by selecting the least cost between multiple unicast routing} $C_t$ \textup{and dual-path routing} $C_p$ \textup{as} $C_i = min (C_t,C_p)$, \textup{where}
		\\
		$C_t = \sum_{k=0}^{|V_i|-1}|d_{kx}-r_x|+|d_{ky}-r_y|$, $|V_i|$ \textup{gives the number of destinations in} $V_i$,\\
		$C_p$ = $Cost_{D_L} + Cost_{D_H}$  \textup{i.e.,the hop count of dual-path \cite{b6}} \textup{from node} $R$.
	\end{definition}
	
	\begin{definition} \label{definition-saving}
		\textup{The} $saving$ $A$ \textup{for merging} t \textup{number of partitions, i.e., for partitions in} $\mathbb{P}$ ,\textup{is obtained by the following rule:} $A$ = max($0$ , $\sum_{i=0}^{t-1}C_i$ - $C_{0,1,..,t-1}$), where $C_{0,1,..,t-1}$ \textup{is the cost of the merged partition.}
	\end{definition}

	\begin{algorithm}
		\DontPrintSemicolon
		\SetAlgoLined
		\SetNoFillComment
		
		%\SetSideCommentLeft
		\SetKwInOut{Input}{Input}
		\SetKwInOut{Output}{Output}
		\Input{Multicast destination set $D = {d_0,d_1,..,d_{N-1}}$ and a source node $S=(s_x,s_y)$}
		\Output{Final partition set $I$}

		$P\leftarrow\{P_0,P_1,..,P_7\}$ based on $S$\\
		$\mathbb{P}\leftarrow\{P_0P_1,P_1P_2,..,P_7P_0,P_0P_1P_2,P_1P_2P_3,..,P_7P_0P_1\}$\\
		$V\leftarrow P\cup \mathbb{P}$
		
		\For {$\forall V_i \in V$ and\ $V_i\ \neq \emptyset$}
		{find $R$ \tcp*{ Definition \ref{definition-representative}}
			\tcc{saving for $P_i \in P$ is not needed}
			\If{$V_i \in \mathbb{P}\  $}
			{find $A(V_i)$ \tcp*{ Definition \ref{definition-saving}}
			}
			
		}
		
		$I$ $\leftarrow$ $\emptyset$ \tcp*{initialize the final partition $I$}
		\While{$\exists A(V_i) \in \mathbb{P} \neq 0$}
		{
			$q \leftarrow \underset{i}{arg\ max} \ \{A(V_i) : \forall \ V_i \ \epsilon \ \mathbb{P}\}$\\ 
			
			$I$ $\leftarrow$ $I \cup V_q$ \\
			
			\tcc{to avoid duplicate delivery}
			\For {$\forall V_i \in V$ and\ $V_q\ \cap V_i \neq \emptyset$}
			{$A(V_i) \leftarrow 0$ }

		}
		\tcc{add leftover partition which did not merge}
		$P \leftarrow P - (P \cap I)$\\
		$I$ $\leftarrow$ $I \cup P$ \\

		\vspace{-0.1cm}
		\caption{Dynamic Partition Algorithm (DPM)}
		\label{algorithm-dynamic}
	\end{algorithm}

	The proposed DPM algorithm is shown in Algorithm \ref{algorithm-dynamic}. For each node at $(x,y)$ in a nxn mesh, it is labeled as $L(x,y) = yn+x$ if y is even, and $L(x,y)=yn+n-x-1$ if y is odd. The labels will be used for calculating the dual-path routing paths. The network is first partitioned in $P$ according to the partition rule, then the extended partition set $\mathbb{P}$ and the search set $V$ are obtained. For each partition in $V$, a representative node $R$ is found, followed by the saving $A$ for each element in $\mathbb{P}$ is calculated. The partition in $\mathbb{P}$ with the maximum saving will be selected as a final partition in the set $I$. The saving $A$ of the other partitions that contain the selected partition will be reset to $0$ to avoid overlapping of the selected partitions. The searching process will quickly converge (up to 4 iterations) to a point when there is no more saving. Any non-merged partition from $P$ is added to $I$. For each partition in the final set $I$, a multicast packet is formed with the destination group as well as the selected routing algorithm and delivered to the representative node $R$ using $XY$ routing. From the representative node $R$, the selected routing algorithm (either  dual-path or multiple unicast based on the comparison in Definition \ref{definition-treevspath}) will be used to deliver the packet.
	
\begin{figure*}
		\centering
		\includegraphics[width=0.85\textwidth]{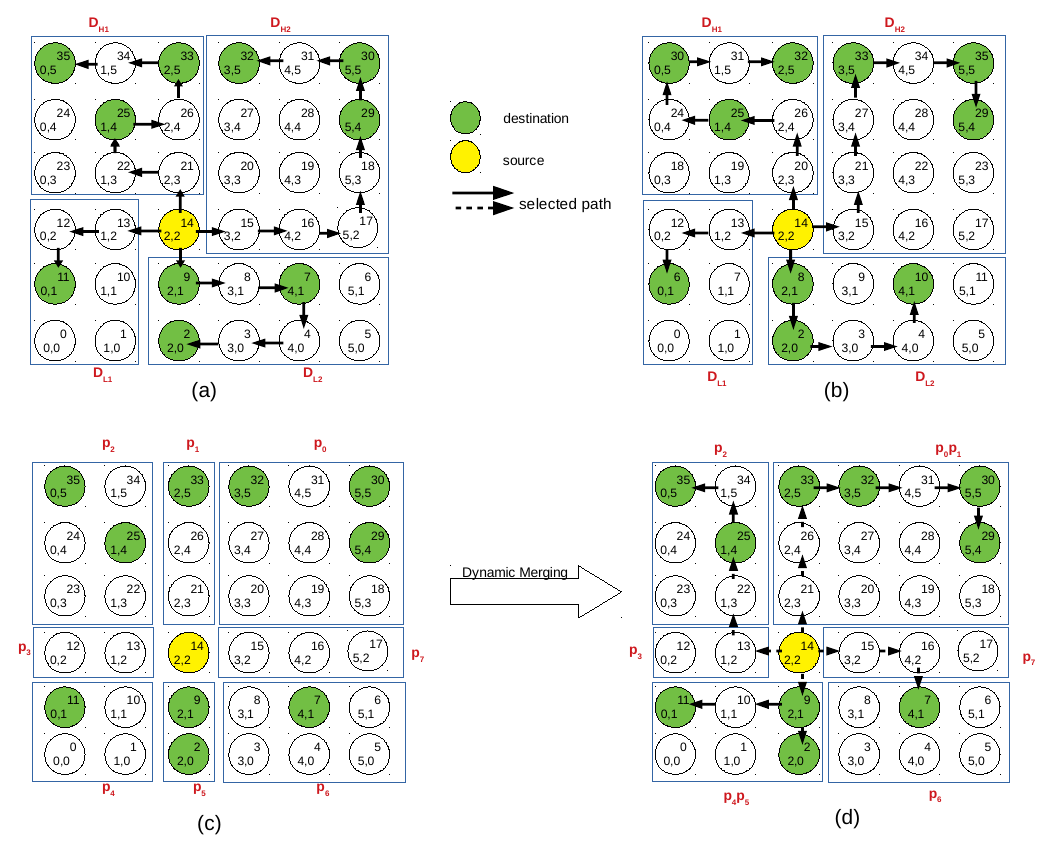}
		\caption{Comparison of (a) MP \cite{b6} (b) NMP \cite{b11} (c) Partition of the network (d) DPM}
		\label{fig-comparision}
	\end{figure*}

	Fig. \ref{fig-comparision} shows the comparison of the multipath (MP) routing\cite{b6}, the new multipath (NMP) routing proposed by Ebrahimi et.al \cite{b11} and the proposed (DPM)-based routing algorithm. In the example, the source node is marked in yellow and all destination nodes are marked in green. Fig. \ref{fig-comparision}a shows the operation of MP, the destination set is divided into four partitions, which are $D_{H1} = \{25,33,35\}$ , $D_{H2} = \{29,30,32\}$, $D_{L1} = \{11\}$, and $D_{L2} = \{9,7,2\}$. For $D_H$ partition, the destination nodes are sorted in the ascending order of node labels and delivered accordingly. Whereas in the $D_L$ partition, destinations are sorted in the descending order of node labels and delivered accordingly. Fig. \ref{fig-comparision}b shows the operation of NMP, where the partition is similar to that of MP, however, the node is label as $L(x,y) = yn+x$. In each partition, the destination is sorted according to the least hop from the source node. A multicast packet is forwarded to the nearest destination first and then again sorted for the next destination. This method is better than multipath in saving the number of routing hops in $D_{H2}$ and $D_{L2}$.
	
	Fig. \ref{fig-comparision}c shows the partition before merging and Fig. \ref{fig-comparision}d shows the partition after merging where $P_0,P_1$ and $P_4,P_5$ are merged. If the cost of merging is same then the group containing the least number of partitions is chosen first then the partition with the smallest indexed partition is chosen. For partition $P_0P_1$, after delivery to the representative node $33$ (a dotted path), a dual-path routing is performed for the rest of the destinations. For partition $P_2$, a unicast is performed to deliver a packet to destination $35$. In partition $P_4P_5$, after delivery to the representative node $9$, multiple unicast (MU) option is chosen as both dual-path and MU gives the same cost but the overhead of computing $D_H, D_L$ is eliminated using MU. The proposed method uses dynamic partition to regroup the eight original partitions into four partition as shown in Fig. \ref{fig-comparision}d. With this optimization, DPM is able to deliver with less number of hops than MP \cite{b6} or NMP \cite{b11}. 
	
	\begin{figure}
		\centering
		\includegraphics[width=\linewidth]{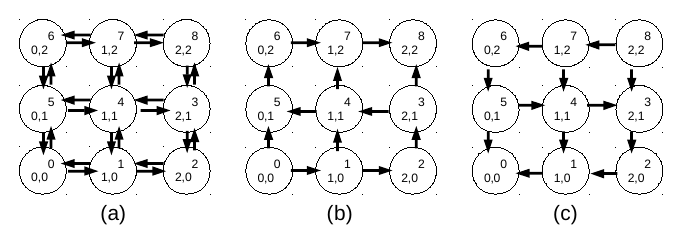}
		\caption{(a) 3x3 mesh physical network (b) high-channel subnetwork (c) low-channel subnetwork}
		\label{fig-channel}
	\end{figure}

	\subsection{Deadlock Avoidance}
	Deadlock is an important aspect to consider when designing a multicast routing algorithm. The proposed DPM-based multicast routing algorithm is likely to produce a deadlock because of mixed turns. To avoid deadlocks, the physical network is divided into high-channel and low-channel subnetwork as shown in Fig. \ref{fig-channel} where the high-channel subnetwork is used when the next hop label is higher than the present node label and the low-channel subnetwork is used if the next hop label is lower than the present node label. Fig. \ref{fig-channel} also shows the allowed turns in each subnetwork, invalid turns are not allowed. As long as this rule is followed deadlocks can be avoided. This method of channel selection and restriction of turns are used by both multicast and unicast traffic to avoid deadlocks.

	\section{Performance Evaluation} \label{experiment}
	In the following sections, the proposed DPM-based multicast routing algorithm is evaluated against the multipath (MP) \cite{b6} and the new multipath (NMP) \cite{b11} algorithms using synthetic traffic and real application workloads.
	
	\subsection{Simulation Settings} \label{simsetting}
\begin{figure}
		\centering
		\includegraphics[width=\linewidth]{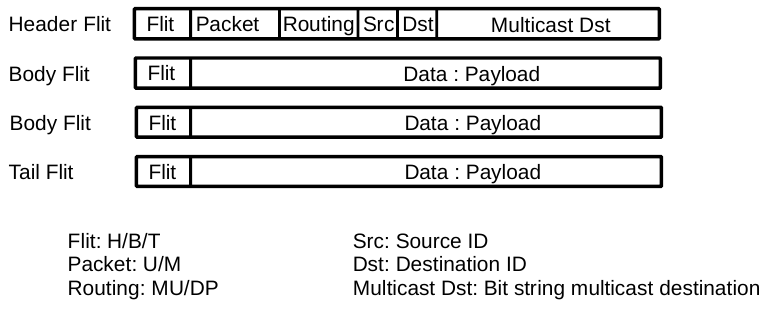}
		\caption{Packet format}
		\label{fig-packet}
	\end{figure}

	The C++ based cycle accurate NoC simulator \cite{b16} is used to simulate the proposed multicast routing algorithm with the network parameter shown in Table \ref{sim param}. Orion 2.0 \cite{b17} is used to estimate the dynamic power consumption. The 8x8 mesh based NoC is used for the simulation purpose where 4 VCs are used, 2 VCs are used for the high and low channel subnetwork respectively. Different destination ranges are adopted based on the study of average destination set size per multicast packet\cite{b18}. In our simulations, on average 10\% of multicast portion and 90\% unicast portion are assumed for the synthetic traffic. Netrace \cite{b19} is used to generate a trace file of workloads from PARSEC \cite{b20} benchmark. Netrace uses 64 cores, in-order Alpha ISA 2Ghz as the target system with 32KB I/D cache with MESI  coherence protocol for L1 cache. The packet format for the simulation is shown in Fig. \ref{fig-packet} where the flit field identifies a head, body or tail flit. The packet field can identify a multicast or unicast packet. The routing field indicates the dual-path or multiple unicast routing algorithm, as determined in Definition \ref{definition-treevspath}. In the header flit, a source node ID, a destination ID followed by multicast destination in bit string format are included. A body or tail flit only contains the flit type identifier and the payload.  
	
	\begin{table}
		\begin{center}
			\caption{Network Configuration}
			\label{sim param}
			\resizebox{0.7 \columnwidth}{!}{
				\begin{tabular}{|c|r|}
					\hline
					Topology & 8x8 Mesh\\
					\hline
					Virtual Channels & 4\\
					\hline
					Buffer Depth & 4 flits\\
					\hline
					Packet Size & 4 flits/packet\\
					\hline
					Traffic  &Uniform Random\\
					\hline
					Multicast Packet Portion & 10 \%\\
					\hline
					Multicast Destination Range & 2-5,4-8,7-10,10-16\\
					\hline		
			\end{tabular}}
		\end{center}
	\end{table}
	
	\begin{figure}
		\centering
		\subfloat[destination range (2-5)]{\includegraphics[width=0.5\textwidth]{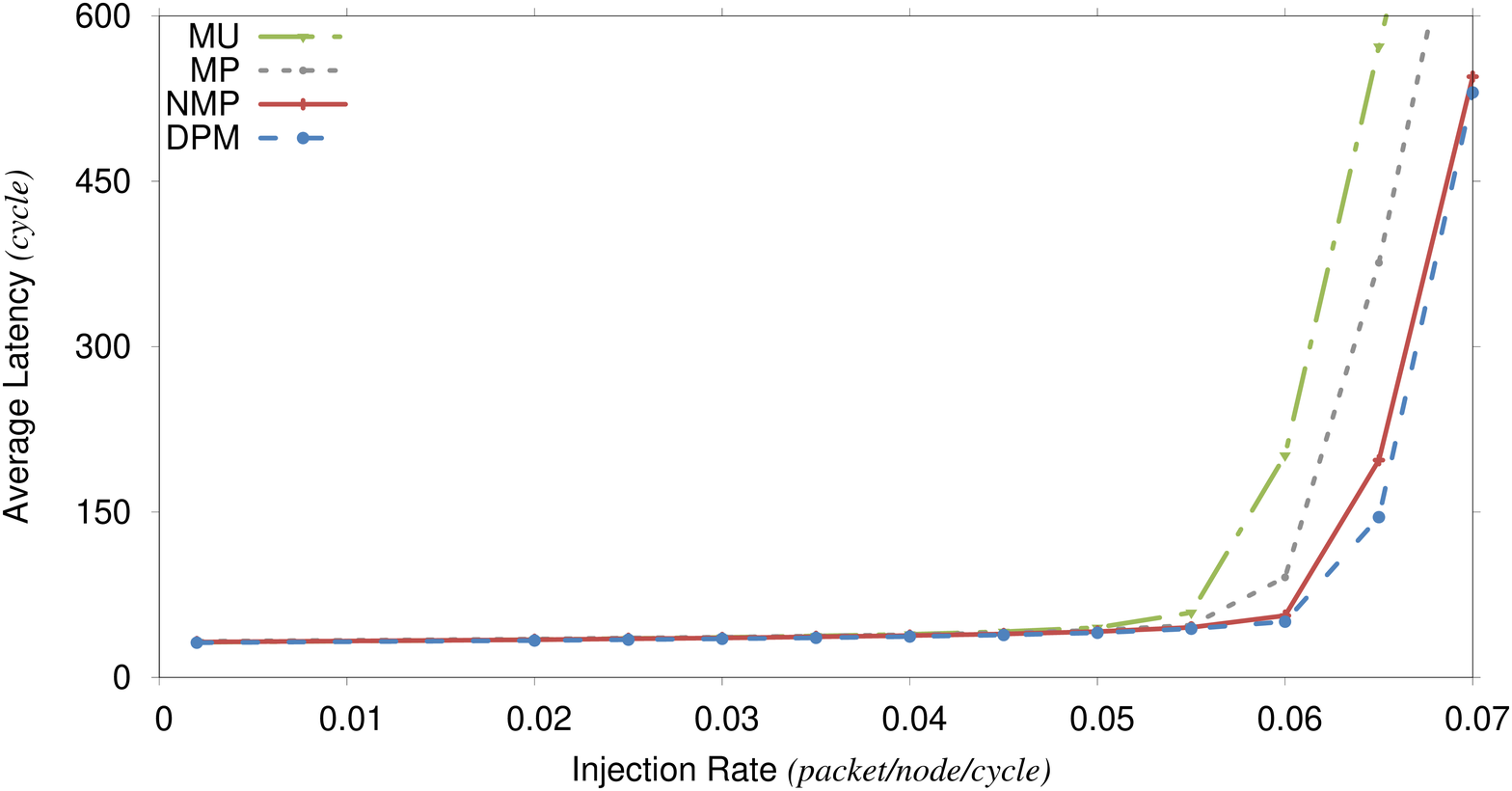}}
		
		\subfloat[destination range (4-8)]{\includegraphics[width=0.5\textwidth]{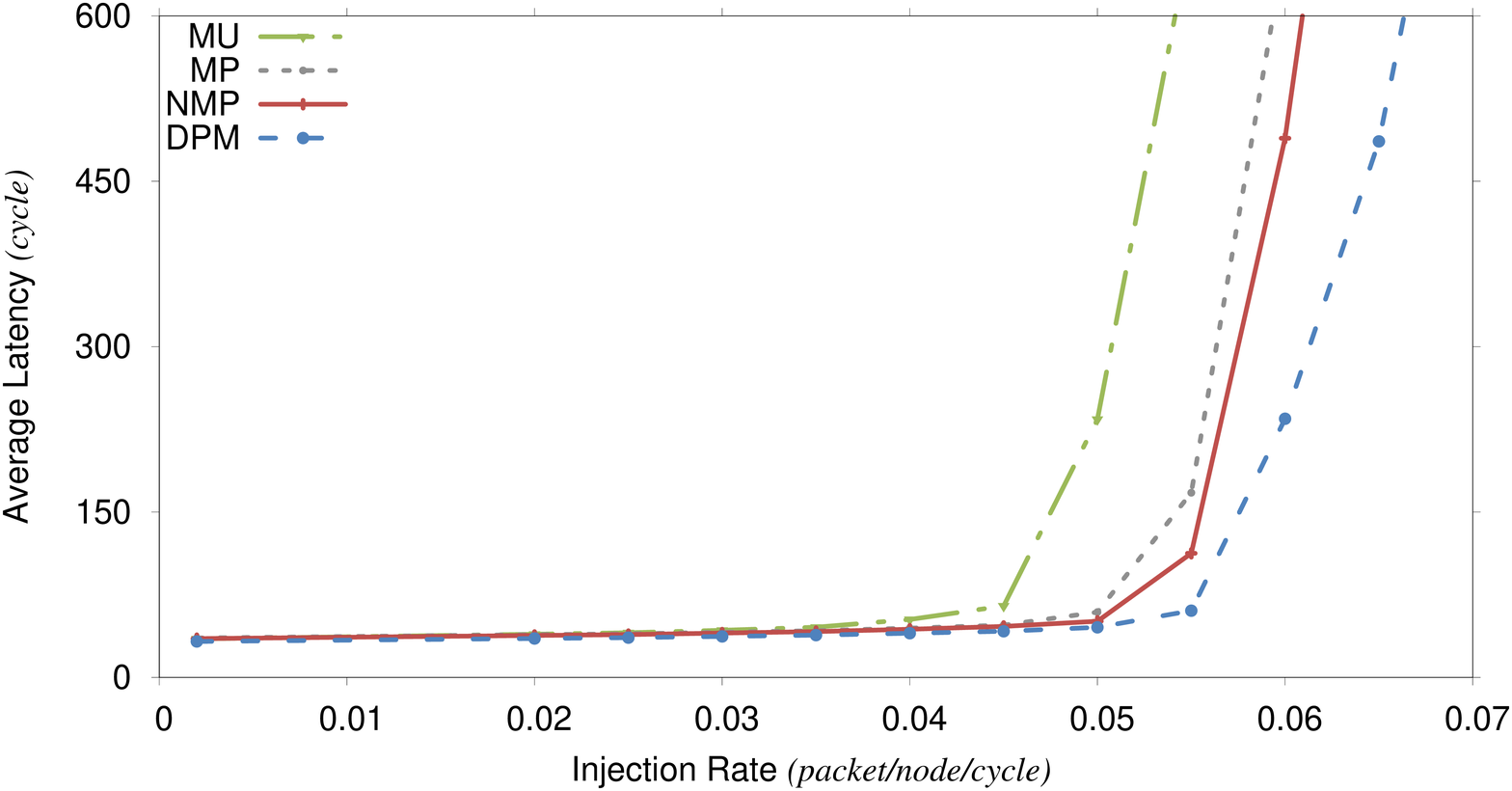}}
		
		\subfloat[destination range (7-10)]{\includegraphics[width=0.5\textwidth]{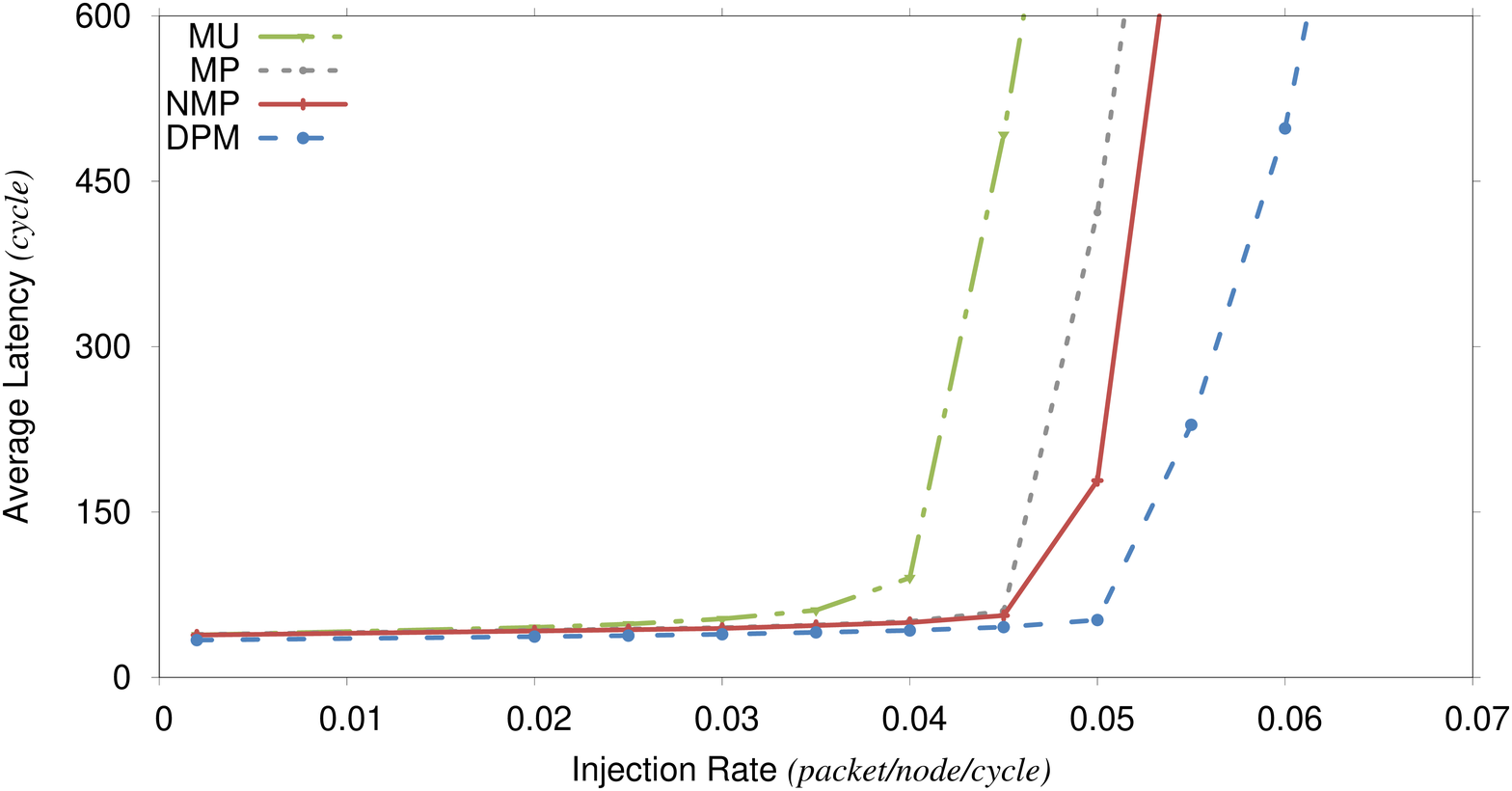}}
		
		\subfloat[destination range (10-16)]{\includegraphics[width=0.5\textwidth]{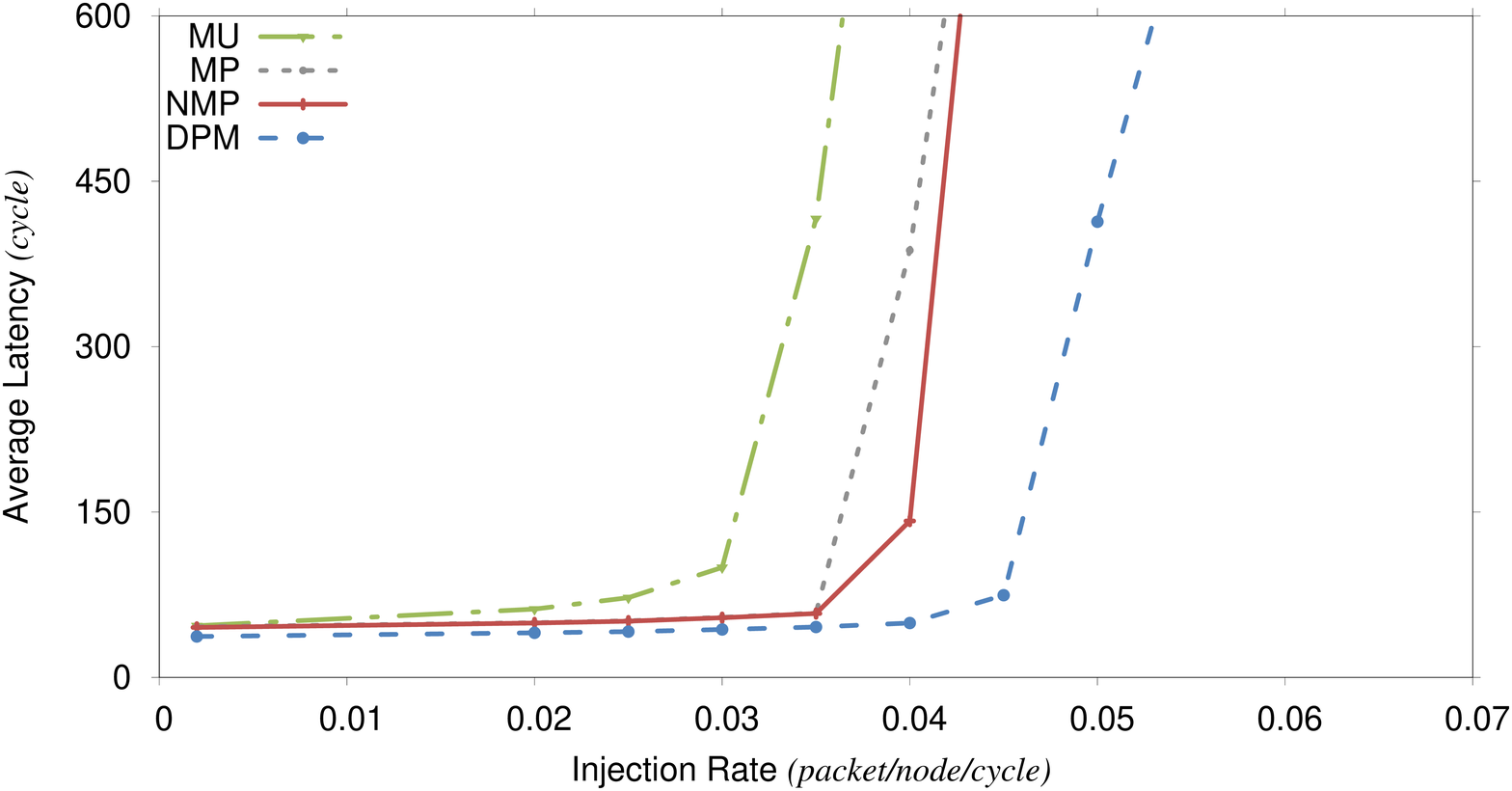}}
		\caption{Average latency for different multicast destination range}
		\label{fig:syntheticlatency}
	\end{figure}

	\subsection{Result} \label{perfcomparison}
	Fig. \ref{fig:syntheticlatency} shows the simulation results for the average packet latency of 8x8 mesh network with synthetic traffic. The multiple unicast (MU) where a multicast packet is delivered by multiple unicast packets, MP \cite{b6}, NMP \cite{b11} and DPM are compared. As the destination range increasing from $(2-5)$ to $(10-16)$, all the routing algorithms are saturated earlier mainly because of the increased congestion in the network. The proposed algorithm has better average packet latency in all destination ranges. The improvement is mainly due to the dynamic partition merging which allows the regrouping of destinations to reduce the routing cost, i.e., saving the hops to deliver leads to comparatively less network congestion.
	
\begin{figure}
		\centering
		\includegraphics[width=0.5\textwidth]{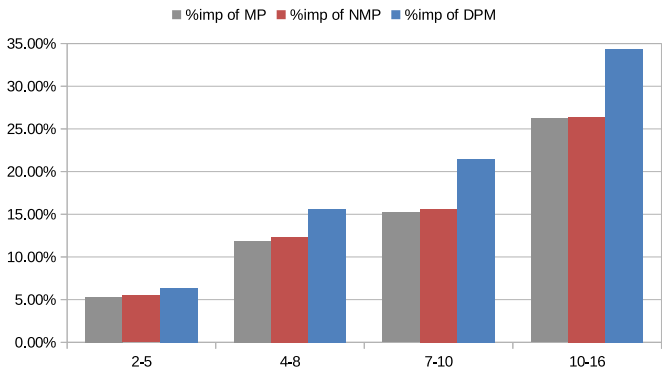}
		
		\caption{\% improvement in power for MP, NMP and DPM over MU}
		\label{fig:syntheticpower}
	\end{figure}

	The dynamic power consumption for all four algorithms are measured for synthetic traffic and the percentage improvement of the latter three algorithms vs. MU at the saturation point of the MU is shown in Fig. \ref{fig:syntheticpower}. DPM is consuming around $7\%,16\%,22\%$, and $35\%$ less power than the MU algorithm at the saturation point for the destination range $(2-5),(4-8),(7-10)$ and $(10-16)$, respectively. For the other two algorithms as well we see a similar trend of the increasing improvement in power consumption up to around $25\%$ improvement at the range $(10-16)$. Similarly, on average DPM is consuming around $21\%$ less power than NMP and $23\%$ less power than MP for all four destination range. 
	
\begin{figure}
		\centering
		\subfloat[Average Packet Latency]{\includegraphics[width=0.5\textwidth]{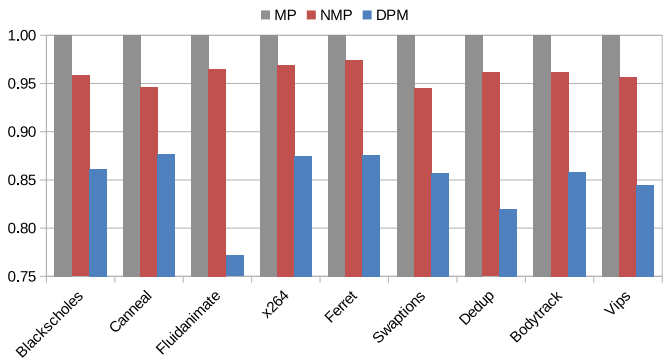}}
		
		\subfloat[Total Power]{\includegraphics[width=0.5\textwidth]{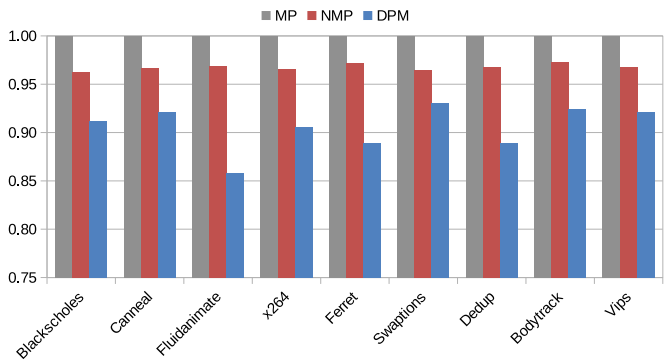}}
		\caption{Performance under PARSEC benchmark traffic}
		\label{fig:benchmark}
	\end{figure}

	The performance of average packet latency under the PARSEC benchmark traces is shown in Fig. \ref{fig:benchmark} where multipath \cite{b6} is considered as a baseline for the benchmark evaluation. For each benchmark, the trace file of the region of interest section is extracted for the simulation. The performance improvement of NMP and DPM vs. MP is  measured. For the average latency in Fig. \ref{fig:benchmark}a, the proposed algorithm shows around $23\%$ improvement in fluidanimate and NMP shows around $5\%$ improvement in canneal and swaptions workloads. Similarly, for the total power consumption, the proposed algorithm shows around $14\%$ improvement in fluidanimate while NMP shows around $4\%$ improvement in swaptions and blackscholes. The variation in delay and power performance in benchmark traces is due to the variation in the destination size and multicast percentage in the benchmark traces. 

	\section{Conclusion} \label{conclusion}
	In this paper, the dynamic partition merging (DPM)-based multicast routing algorithm is proposed for NoCs. The destination set partition problem is first modeled as the exact weighted set cover problem. The proposed DPM algorithm is proposed to solve the problem by comparing the different options of merging partitions and selecting the option with the most saving in routing cost. Compared with the static partition strategy, DPM finds the partitions with lower routing cost and hence helps in reducing the network latency and power consumption. The routing algorithm inside each partition is selected from the dual-path and multiple unicast routing. The simulation results show that for synthetic traffic, the proposed algorithm achieves lower average packet latency and the network is saturated later than the existing algorithms with less power consumption. Similarly, for PARSEC benchmark loads, latency improvement is up to 23\% and power improvement is up to 14\% are observed. In our future work, the extension of the DPM-based multicast routing algorithm to 3D NoCs and evaluation of machine-learning benchmarks will be conducted.

\end{document}